\documentclass[9pt]{article}
\usepackage{spconf,amsmath,graphicx}
\usepackage{amssymb}
\usepackage{booktabs}
\usepackage{multirow}
\usepackage[english]{babel}
\usepackage{amsthm}
\usepackage{setspace}
\usepackage{cite}
\usepackage{algorithm,algorithmic}
\usepackage[normalem]{ulem}
\useunder{\uline}{\ul}{}
\usepackage{url}

\def\Mat#1{\boldsymbol{\mathbf{#1}}}

\def\Vec#1{\textsf{\boldmath $#1$}}

\def\E{{\mathbb E}}

\def\R{{\mathbb R}}
\def\T{{\textsf{T}}}

\def\thline{\noalign{\hrule height 1.2pt}}

\newcommand{\refeq}[1]{(\ref{eq:#1})}

\newcommand{\reffig}[1]{Fig. \ref{fig:#1}}

\newcommand{\reftab}[1]{Table \ref{tab:#1}}

\title{Remixed2Remixed: Domain adaptation for speech enhancement \\
by Noise2Noise learning with Remixing}
\name{Li Li, Shogo Seki
\thanks{Accepted by IEEE ICASSP 2024.
© 2024 IEEE. Personal use of this material is permitted. Permission from IEEE must be obtained for all other uses, in any current or future media, including reprinting/republishing this material for advertising or promotional purposes, creating new collective works, for resale or redistribution to servers or lists, or reuse of any copyrighted component of this work in other works.}
}
\address{CyberAgent, Inc.}

\begin{document}

\maketitle

\begin{abstract}
This paper proposes Remixed2Remixed, a domain adaptation method for speech enhancement, which adopts Noise2Noise (N2N) learning to adapt models trained on artificially generated (out-of-domain: OOD) noisy-clean pair data to better separate real-world recorded (in-domain) noisy data. 
The proposed method uses a teacher model trained on OOD data to acquire pseudo-in-domain speech and noise signals, which are shuffled and remixed twice in each batch to generate two bootstrapped mixtures. The student model is then trained by optimizing an N2N-based cost function computed using these two bootstrapped mixtures.
As the training strategy is similar to the recently proposed RemixIT, we also investigate the effectiveness of N2N-based loss as a regularization of RemixIT. 
Experimental results on the CHiME-7 unsupervised domain adaptation for conversational speech enhancement (UDASE) task revealed that the proposed method outperformed the challenge baseline system, RemixIT, and reduced the blurring of performance caused by teacher models.
\end{abstract}

\section{Introduction}
Speech enhancement (SE) \cite{Loizou2007} is one of the fundamental problems in speech signal processing and serves many applications, either as a hearing aid or as a frontend system for many other tasks.
The goal is to improve speech quality recorded in the presence of noise, interference, and reverberation, which has been greatly advanced by deep neural networks (DNNs).
\par
Supervised learning is the most studied approach to SE \cite{Ochieng2022deep}, in which the model is trained on noisy-clean pair data to predict clean signals directly \cite{Macartney2018improved, Défossez2020real} or by masking \cite{Luo2019conv, Tzinis2020sudo, Zhao2021monaural}. 
Since recording such parallel pair data is impossible due to crosstalk \cite{Ito2023audio}, artificially synthesized noisy data is generally used to train SE models. 
However, due to the distribution mismatch mainly caused by the different acoustic conditions between such synthetic (out-of-domain: OOD) and real-world recorded (in-domain) data, the trained models usually suffer from performance degradation when faced with recorded data. 
Several methods have been proposed recently to address this issue, including unsupervised methods aimed at learning models on nonparallel data. 
This can be achieved, for example, by using machine learning methods that learn from positive and unlabeled data \cite{Ito2023audio}, by replacing the ground truth of clean speech with evaluation metric scores \cite{Subramanian2019speech, Fu2022metric}, and by using observation consistency \cite{Wisdom2020unsupervised, Saijo2021self}.
\par
Another effective solution is to perform domain adaptation, which adjusts an SE model pre-trained on OOD data to formulate an accurate noisy-clean mapping that matches in-domain data. 
Existing methods include using adaptive mechanisms such as adversarial learning, optimal transport \cite{Liao2019noise, Lin2021unsupervised}, and self-supervised learning. 
RemixIT \cite{Tzinis2022remixit} is one method using self-distillation, which consists of two networks. A teacher model pre-trained with synthesized OOD pair data\footnote{RemixIT can be trained in a fully unsupervised manner, where the teacher model is trained solely with noisy speech by MixIT \cite{Wisdom2020unsupervised}.} is used 
to produce the pseudo-paired data of noisy speech and target signals for student training by remixing separated speech and noise signals in each batch. A student model is then trained using the generated pseudo-paired data by minimizing the loss between the predicted signals and pseudo-targets. 
The teacher model is continually updated with a weighted moving average (WMA) using the student model's weights.
Although RemixIT loss has been theoretically shown to ideally approach supervised loss when the teacher model can accurately predict signals or when the student model can see many pseudo-mixtures containing the same teacher estimates, it is not feasible with limited training resources. 
As a result, the performance of RemixIT depends to some extent on the performance of its teacher model.
\par
On the other hand, approaches that apply basic statistical reasoning have been proposed for DNN-based image denoising.
Based on the principle that corrupting the training target of the network with zero-mean noise does not change what the denoising network learns from the clean signal, Noise2Noise (N2N) \cite{Lehtinen2018noise} demonstrated that a denoising model could be trained on noisy-noisy pair data, which was later extended to SE \cite{Kashyap2021speech}. 
However, it is still difficult to collect paired data containing two independent noisy realizations of the same clean signal, especially for audio signals. 
This motivates more improved methods to remove demands on data further. Noisier2Noise (Nr2N) \cite{Moran2020noisier} and recorrupted-to-recorrupted (R2R) \cite{Pang2021recorrupted} use noise sampled from a known prior distribution to generate noisy pair data for image denoising. Noisy-target training (NyTT) \cite{Fujimura2021noisy, Sivaraman2021personalized} uses noisy speech with additional noise to obtain noisy pair data for SE.
It has shown that NyTT can reduce noise close to the additional noise used in training, but performance degrades when faced with other noise \cite{Fujimura2023analysis}.
\par
Considering the potential of learning models with less in-domain data than unsupervised learning that learns from scratch, this paper focuses on the domain adaptation approach and proposes a method called {\it Remixed2Remixed (Re2Re)}, which employs a teacher-student architecture similar to RemixIT and N2N learning. 
Specifically, the teacher model is used to generate pseudo-noisy pair data by performing the remix procedure twice, and the student model is trained using an N2N-based cost function. 
This allows both in-domain speech and noise to be obtained from noisy speech only. 
Moreover, by explicitly optimizing a cost function defined for denoising, the proposed method is expected to perform more consistently than RemixIT, regardless of the performance of the teacher model. 

\section{Conventional method: RemixIT}
\subsection{Supervised learning}
Let us denote speech and noise signals drawn from corresponding distributions by  
$\Vec{s}\sim\mathcal{D}_{\Vec{s}}$ and $\Vec{n}\sim\mathcal{D}_{\Vec{n}}$, respectively. 
Synthetic noisy speech can be obtained as $\Vec{x}=\Vec{s}+\Vec{n}$.
With pair data $(\Vec{x}, \Vec{s}, \Vec{n})$, 
a model predicting both speech and noise $\hat{\Vec{s}}, \hat{\Vec{n}} = \mathcal{F}(\Vec{x}; \theta)$ parameterized by $\theta$ can be trained under full supervision by optimizing the following cost function (i.e., minimizing the reconstruction error of both signals):
\begin{align}
\mathcal{L}_{\rm supervised}=\E_{(\Vec{x}, \Vec{s}, \Vec{n})}\big[\mathcal{L}(\hat{\Vec{s}}, \Vec{s}) + \mathcal{L}(\hat{\Vec{n}}, \Vec{n})\big].
\end{align}

\subsection{RemixIT}
RemixIT \cite{Tzinis2022remixit} comprises a teacher model $\mathcal{F}_{\mathcal{T}}$ and a student model $\mathcal{F}_{\mathcal{S}}$. Both models are initialized with a supervised pre-trained model using synthetic OOD pair data $(\Vec{x}, \Vec{s}, \Vec{n})$ and further trained to enhance better real-world recorded data $\Vec{x}'\sim \mathcal{D}_{\Vec{x}'}$ with only the in-domain data accessible.
Given a mini-batch of in-domain noisy data $\Mat{x}'=\Mat{s}'+\Mat{n}'\in\R^{B\times T}$, the teacher model estimates speech and noise signals as follows
\begin{align}
\tilde{\Mat{s}}', \tilde{\Mat{n}}' = \mathcal{F}_{\mathcal{T}}(\Mat{x}'; \theta_{\mathcal{T}}^{(k)}),
\end{align}
where the bold roman font represents a batch $\Mat{a}=[\Vec{a}_1, \ldots, \Vec{a}_B]^\T$ including multiple signals $\Vec{a}_b$ drawn from distribution $\mathcal{D}_{\Vec{a}}$ and $\theta_\mathcal{T}^{(k)}$ denotes the parameters of teacher model at the $k$-th training epoch. $^\T$ denotes transpose operator and $B$ and $T$ denote mini-batch size and signal length, respectively. 
The estimated signals are then shuffled and remixed to generate bootstrapped mixture $\tilde{\Mat{x}}'$, which is expressed as
\begin{align}
\tilde{\Mat{x}}'=\tilde{\Mat{s}}' + \Mat{P}\tilde{\Mat{n}}'.
\label{eq:bootstrapped1}
\end{align}
Here, $\Mat{P}\sim \Pi_{B\times B}$ is a permutation matrix.
The bootstrapped mixture is then used to generate in-domain pseudo-paired data $(\tilde{\Mat{x}}', \tilde{\Mat{s}}', \tilde{\Mat{n}}')$.
The student model $\mathcal{F}_{\mathcal{S}}$ with parameter $\theta_\mathcal{S}^{(k)}$ is then trained by minimizing the reconstructed error between the outputs of the model and the pseudo-targets $\tilde{\Mat{s}}'$ and $\tilde{\Mat{n}}'$ as follows:
\begin{align}
\hat{\Mat{s}}', \hat{\Mat{n}}' &= \mathcal{F}_{\mathcal{S}}(\tilde{\Mat{x}}'; \theta_{\mathcal{S}}^{(k)}), \\
\mathcal{L}_{\rm RemixIT} &= \sum_{b=1}^B \big[
\mathcal{L}(\hat{\Mat{s}}'_b, \tilde{\Mat{s}}'_b) + \mathcal{L}(\hat{\Mat{n}}'_b, [\Mat{P}\tilde{\Mat{n}}']_b) \big].
\label{eq:RemixIT}
\end{align}
To generate more accurate pseudo-targets, the teacher model is continuously updated by taking a weighted moving average (WMA) with the student model's weights at constant epoch, which is expressed by
$\theta_{\mathcal{T}}^{(k+1)} = \gamma \theta_{\mathcal{S}}^{(k)} + (1-\gamma) \theta_{\mathcal{T}}^{(k)}$. 
Here, $0\leq\gamma\leq 1$ is a weight parameter.
\par
It is noteworthy that the cost function of RemixIT $\mathcal{L}_{\rm RemixIT}$ has convergence properties when Euclidean
norm-based metric is used to measure the reconstruction error:
\begin{align}
&\mathcal{L}_{\rm RemixIT} \propto 
\E \big[||\hat{\Mat{s}}'-\tilde{\Mat{s}}'||^2_2\big] 
=\E \big[||(\hat{\Mat{s}}'-\Mat{s}')-(\tilde{\Mat{s}}'-\Mat{s}')||^2_2\big] \nonumber \\
=&\E \big[||(\hat{\Mat{s}}'-\Mat{s}')||^2_2\big] + \E \big[||(\tilde{\Mat{s}}'-\Mat{s}')||^2_2\big] - 2\E\big[(\tilde{\Mat{s}}'-\Mat{s}')^\T(\hat{\Mat{s}}'-\Mat{s}')\big] \nonumber \\
\approx &\underbrace{\E \Big[||\Mat{\epsilon}'_\mathcal{S}||^2_2\Big]}_{\rlap{\rm \footnotesize supervised~loss}} - \underbrace{\E \Big[||\Mat{\epsilon}'_\mathcal{T}||^2_2\Big]}_{\rlap{\rm \footnotesize constant~w.r.t~$\theta_\mathcal{S}$}} 
-2\E\Big[\underbrace{(\tilde{\Mat{s}}'-\Mat{s}')^\T}_{\rlap{\rm \footnotesize \hspace{-10pt}teacher~error}}\underbrace{\frac{1}{M}\sum_{m=1}^M(\hat{\Mat{s}}'_m-\tilde{\Mat{s}}')}_{\rlap{\rm \footnotesize \hspace{-15 pt}empirical~mean~student~error}}\Big], 
\label{eq:remixit_mse}
\end{align}
where $\Mat{\epsilon}'_\mathcal{S}$ and $\Mat{\epsilon}'_\mathcal{T}$ are reconstruction errors between the target signal $\Mat{s}'$ and the output of the student and teacher models, respectively, and $||\cdot||_2^2$ denotes squared $L2$ norm.
\refeq{remixit_mse} shows that when the third term is zero, the RemixIT loss approaches the supervised loss. 
This can be achieved by reducing either the teacher error to zero with an accurately estimated signal in the teacher model or the empirical mean student error to zero by exposing the student to a wide variety of bootstrapped mixtures $\tilde{\Mat{x}}'_m=\tilde{\Mat{s}}'+ \tilde{\Mat{n}}'_m,~m=1,\ldots, M$ involving the same teacher estimate $\tilde{\Mat{s}}'$
so that $\E_m[\hat{\Mat{s}}'_m|\tilde{\Mat{x}}'_m]=\tilde{\Mat{s}}'$ when $M\to\infty$.
This property is important to ensure that RemixIT can learn models as supervised learning. 
However, reducing the third term to zero with limited training resources, for example, with $M=1$ is not feasible. As a result,  the performance of RemixIT inevitably depends to some extent on the performance of its teacher model, and furthermore, there may remain a gap with supervised learning.

\section{Proposed method: Remixed2Remixed}
\begin{figure*}[]
\centering
\includegraphics[width=\linewidth]{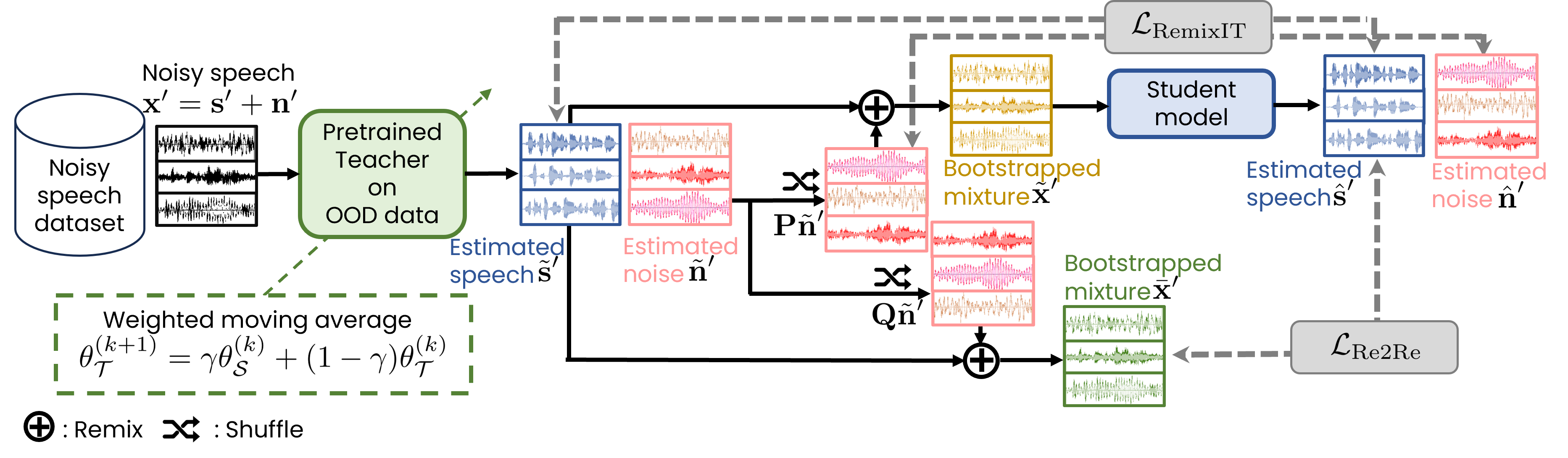}
\caption{Flowchart of proposed Remixed2Remixed.}
\label{fig:flowchart}
\end{figure*}

N2N \cite{Lehtinen2018noise} is an image denoising method utilizing basic statistical reasoning, which has demonstrated that a denoising model could be trained using noisy pair data $(\Vec{x}, \bar{\Vec{x}})$ instead of $(\Vec{x}, \Vec{s})$ if the noisy signal $\bar{\Vec{x}}=\Vec{s}+\bar{\Vec{n}}$ satisfies $\E[\bar{\Vec{x}}|\Vec{x}]=\Vec{s}$. This can be achieved when $\E[\bar{\Vec{n}}]=0$ and $\bar{\Vec{n}}$ and $\Vec{n}$ are independent to each other, namely, $\Vec{x}$ and $\bar{\Vec{x}}$ are two independent noisy realizations of $\Vec{s}$.
Inspired by the success of N2N, we would like to extend it to SE with motivation similar to that of \cite{Kashyap2021speech}. 
Different from \cite{Kashyap2021speech}, where paired data of two noisy realizations is obtained synthetically, we utilize the teacher-student architecture in RemixIT to generate paired noisy data via remixing in-domain speech and noise signals separated by a pre-trained OOD model. 
This makes it easy to obtain two in-domain noisy realizations containing the same signals from the recorded noisy signal only. 
\par
\reffig{flowchart} demonstrates a flowchart of the proposed method, {\it Remixed2Remixed (Re2Re)}.
Re2Re has a similar teacher-student architecture to RemixIT, with the difference that it generates in-domain paired data of two noisy realizations $(\bar{\Mat{x}}', \tilde{\Mat{x}}')$ by performing the remixing process twice to generate two bootstrapped mixtures every training iteration. 
Besides bootstrapped mixture $\tilde{\Mat{x}}'$ generated using \refeq{bootstrapped1}, another bootstrapped mixture containing teacher estimate $\tilde{\Mat{s}}'$ is given by
\begin{align}
\bar{\Mat{x}}' = \tilde{\Mat{s}}' + \Mat{Q}\tilde{\Mat{n}}' , 
\end{align}
where $\Mat{Q}$ is uniformly sampled from a set of $B\times B$ permutation matrices such that $\Mat{Q}\perp\Mat{P}$.
With noisy pair data $(\bar{\Mat{x}}', \tilde{\Mat{x}}')$, the student model is trained by minimizing an N2N-based loss
\begin{align}
\mathcal{L}_{\rm Re2Re} =  \E_{(\bar{\Mat{x}}',\tilde{\Mat{x}}')}
\big[\mathcal{L}(\mathcal{F}_{\mathcal{S}}(\tilde{\Mat{x}}';\theta_{\mathcal{S}}), \bar{\Mat{x}}')\big]
=
\E_{(\bar{\Mat{x}}',\tilde{\Mat{x}}')}
\big[\mathcal{L}(\hat{\Mat{s}}', \bar{\Mat{x}}')\big],
\label{eq:N2N}
\end{align}
which satisfies $\E[\bar{\Mat{x}}'|\tilde{\Mat{x}}']=\Mat{s}'$ when sufficient paired data $(\bar{\Mat{x}}', \tilde{\Mat{x}}')$ the student model can see.
To generate sufficient pair data, we update teacher model every epoch so that $\tilde{\Mat{x}}'$ and $\bar{\Mat{x}}'$ can be considered as two noisy realizations of signal $\Mat{s}'$ generated in an on-the-fly manner by corrupting speech signal $\Mat{s}'$ with $\Mat{\epsilon}^{(k)}_\mathcal{T}+\Mat{P}\tilde{\Mat{n}}'$ and $\Mat{\epsilon}^{(k)}_\mathcal{T}+\Mat{Q}\tilde{\Mat{n}}'$, respectively. $\Mat{\epsilon}^{(k)}_\mathcal{T}$ is the estimated error of the teacher model in $k$-th epoch. 
It is general to assume noise signals and estimated error are zero means. Therefore, $(\bar{\Mat{x}}', \tilde{\Mat{x}}')$ can be considered to satisfy the zero-mean condition. 
Although $\Mat{\epsilon}^{(k)}_\mathcal{T}+\Mat{P}\tilde{\Mat{n}}'$ and $\Mat{\epsilon}^{(k)}_\mathcal{T}+\Mat{Q}\tilde{\Mat{n}}'$ are not exactly independent due to the presence of $\Mat{\epsilon}^{(k)}_\mathcal{T}$, the impact of $\Mat{\epsilon}^{(k)}_\mathcal{T}$ can be reduced by increasing the power of $\Mat{P}\tilde{\Mat{n}}'$ and $\Mat{Q}\tilde{\Mat{n}}'$. 
We also consider applying the N2N loss as a regularization for RemixIT, referred to as Re2Re\_reg, whose cost function is given by
\begin{align}
\mathcal{L}_{\rm Re2Re\_reg} = 
\mathcal{L}_{\rm RemixIT} + \beta \mathcal{L}_{\rm Re2Re}.
\end{align}
Here, $\beta\geq 0$ is a parameter balancing the importance of each term. 
By explicitly optimizing a cost defined for denoising \refeq{N2N} instead of a reconstruction error \refeq{RemixIT} between outputs of teacher and student models, the methods using N2N loss are expected to perform more consistently than RemixIT, regardless of the performance of the teacher model.

\section{Experimental evaluation}
\label{sec:experiment}
\subsection{Datasets and experimental conditions}
\begin{table}[]
\centering
\caption{SI-SDR [dB] in reverberant CHiME-5 dataset and DNS-MOS in 1-spk subset of CHiME-5 dataset. $^*$ denotes model checkpoints provided by CHiME-7. Other models were trained from Sudo rm-rf$^*$ checkpoints. Bold fonts indicate the best scores.}
\label{tab:result1}
\setlength{\tabcolsep}{0.1pt}
\begin{tabular}{@{\extracolsep{1pt}}lcccccccc@{}}
\thline
\multicolumn{1}{c}{}                          & \multicolumn{4}{l}{CHiME-5 w/o VAD}                                                                                                             & \multicolumn{4}{c}{CHiME-5 w/ VAD}        \\
\cline{2-5}\cline{6-9}
\multicolumn{1}{c}{}                          & \multicolumn{1}{c}{}                                  & \multicolumn{3}{c}{DNS-MOS}                                                             &                                   & \multicolumn{3}{c}{DNS-MOS}                                      \\
\cline{3-5} \cline{7-9} 
\multicolumn{1}{l}{\multirow{-3}{*}{Methods}} & \multicolumn{1}{c}{\multirow{-2}{*}{\begin{tabular}[c]{@{}c@{}}SI-SDR \\ {[}dB{]}\end{tabular}}}
& OVR                         & BAK                         & SIG             & \multicolumn{1}{c}{\multirow{-2}{*}{\begin{tabular}[c]{@{}c@{}}SI-SDR \\ {[}dB{]}\end{tabular}}} & OVR                         & BAK                         & SIG  \\
\hline
Sudo rm-rf$^*$  &  7.80 & {\bf 2.88}  & 3.59  & 3.33 &  7.80 & {\bf 2.88} & 3.59  & {\bf 3.33} \\
RemixIT$^*$ & 9.44 & 2.83  & {\bf 3.65} & 3.25 & 10.05 & 2.84 & {\bf 3.63}  & 3.27 \\
RemixIT     &  10.94 & 2.84 & 3.63 & 3.29 & 10.68   & 2.85 & 3.51 & {\bf 3.33} \\
Re2Re\_reg & 11.26 & 2.82 & 3.54 & 3.31 & 11.64 &2.82 & 3.51 & 3.32 \\
Re2Re & {\bf 11.65} & 2.84 & 3.42 & {\bf 3.37} & {\bf 11.76}  & 2.80  & 3.47 & 3.29 \\
\thline
\end{tabular}
\end{table}

\begin{table*}[th]
\centering
\caption{Average SI-SDRs and standard deviations [dB] over ten trials in reverberant LibriCHiME-5 dataset. All models were initialized by the same teacher models. Bold fonts indicate the best scores, and underlines indicate standard deviations smaller or equivalent to those achieved by RemixIT. }
\label{tab:result2}
\setlength{\tabcolsep}{1pt}
\begin{tabular}{@{\extracolsep{3pt}}lrrrrrrrr@{}}
\thline
                & \multicolumn{4}{c}{CHiME-5 w/o VAD}                                                                                                                                      & \multicolumn{4}{c}{CHiME-5 w/ VAD}                                                                                                                                       \\
                \cline{2-5}\cline{6-9}
\multirow{-2}{*}{Methods} & \multicolumn{1}{c}{1-spk}                 & \multicolumn{1}{c}{2-spk}                 & \multicolumn{1}{c}{3-spk}                 & \multicolumn{1}{c}{Avg.}                  & \multicolumn{1}{c}{1-spk}                 & \multicolumn{1}{c}{2-spk}                 & \multicolumn{1}{c}{3-spk}                 & \multicolumn{1}{c}{Avg.}                  \\
\thline
Sudo rm-rf            & 8.68 $\pm$ 0.63                               & 8.76 $\pm$ 1.02                               & 7.50 $\pm$ 1.55                               & 8.67 $\pm$ 0.75                               & 8.36 $\pm$ 0.86                               & 8.46 $\pm$ 1.15                               & 7.84 $\pm$ 1.43                               & 8.37 $\pm$ 0.95                               \\
RemixIT                   & 10.95 $\pm$ 0.94                              & 10.76 $\pm$ 1.51                              & 9.91 $\pm$ 2.13                               & 10.87 $\pm$ 1.10                              & 11.21 $\pm$ 0.56                              & 11.25 $\pm$ 0.81                              & 10.76 $\pm$ 1.05                              & 11.20 $\pm$ 0.59                              \\
Re2Re\_reg & {\ul \textbf{11.34 $\pm$ 0.48}}	& {\ul 11.20 $\pm$ 0.92} & {\ul 10.53 $\pm$ 1.32}	& {\ul 11.28 $\pm$ 0.57} & {\ul 11.35 $\pm$ 0.46} &	{\ul 11.42 $\pm$ 0.61} &	{\ul 10.84 $\pm$ 0.66}	& {\ul 11.35 $\pm$ 0.48}\\
Re2Re            & {\ul 11.24 $\pm$ 0.39}  & {\ul \textbf{11.75 $\pm$  0.77}} & {\ul \textbf{11.53 $\pm$ 1.19}} &  {\ul \textbf{11.38 $\pm$ 0.45}} &  {\ul \textbf{11.44 $\pm$ 0.49}} & {\ul \textbf{11.83 $\pm$ 0.73}} &	{\ul \textbf{11.61 $\pm$ 0.82}} & {\ul \textbf{11.55 $\pm$ 0.53}} \\
\thline
\end{tabular}
\end{table*}

To evaluate the performance of the proposed Re2Re for domain adaptation, we conducted speech enhancement experiments on the CHiME-7 unsupervised domain
adaptation for conversational speech enhancement (UDASE) task \cite{UDASE, UDASE_web}, which consists of three datasets: (1) the LibriMix paired dataset for training OOD supervised SE model and development; (2) the CHiME-5 in-domain unlabeled dataset for adopting domain adaptation, development, and evaluation; (3) the reverberant LibriCHiME-5 close-to-in-domain paired dataset for development and evaluation. 
All datasets contain three subsets labeled with a maximum number of speakers: 1-spk, 2-spk, and 3-spk.
\textbf{LibriMix \cite{LibriMix}:} A noisy speech separation benchmark comprises clean speech and noise signals from LibriSpeech \cite{LibriSpeech} and WHAM! \cite{WHAM!}, respectively.
Libri2Mix and Libri3Mix with two or three overlapping speakers in each mixture can be used as subsets of 2-spk and 3-spk, and a subset of 1-spk (Libri1Mix) is obtained by discarding one of the two speakers in the Libri2Mix mixtures.
The proportion of 1-spk, 2-spk, and 3-spk mixtures is 0.5, 0.25, and 0.25, respectively.
\textbf{CHiME-5 \cite{CHiME-5}:} A dataset originally consists of noisy multi-speaker speeches of twenty conversation sessions recorded in 4-people dinner parties. 
CHiME-7 UDASE excerpted the recording channel where participants wearing microphones did not speak (i.e., the maximum number of simultaneously active speakers is three) and divided signals into four subsets, including short segments of at least 3 seconds long labeled by the maximum number of speakers according to the transcript.
The subset containing noise-only segments is used to create the reverberant LibriCHiME-5 dataset for objective evaluations. Other subsets are further divided for train ($\approx$83h), development ($\approx$15.5h), and evaluation ($\approx$7h), respectively. 
Segments for training are cut into chunks of up to 10 seconds, and a voice activity detector (VAD) is applied as post-processing to obtain two versions of the training dataset: CHiME-5 w/o VAD and CHiME-5 w/ VAD. 
\textbf{Reverberant LibriCHiME-5:} A synthetic dataset consists of reverberant noisy speech labeled with clean speech, where clean speech and noise signals are excerpted from LibriSpeech \cite{LibriSpeech} and the above-mentioned noise-only subset, respectively. The room impulse responses (RIRs) excerpted from the VoiceHome corpus are recorded in the living room, kitchen, and bedroom of 3 real homes with 18 different microphone arrays and loudspeaker settings. 
The mixtures are generated by adding noise segments into randomly sampled speech utterances convolved with randomly sampled RIRs, where the signal-to-noise ratio (SNR) for each speaker is distributed as a Gaussian with a mean of 5 dB and a standard deviation (std) of 7 dB to match the CHiME-5 dataset. The proportion of 1-spk, 2-spk, and 3-spk subsets was 0.6, 0.35, and 0.05, respectively. 
Data duration for development and evaluation is about 3 hours each.
\par
We used the recipe provided by CHiME-7 without modification except for the cost function to demonstrate the effectiveness of the cost function.
We used Sudo rm-rf \cite{Tzinis2020sudo} architecture for both teacher and student models, whose encoder and decoder consisted of one-dimensional convolution and transpose convolution, respectively, with 512 filters of 41 taps and a hop size of 20 samples, and the separator consisted of 8 U-Conv blocks. 
The pre-trained teacher model initialized the student model and was continually updated by WMA with a weight of $\gamma=0.01$ every epoch. The batch size was 24. Negative scale-invariant signal-to-distortion ratio (SI-SDR) \cite{SI-SDR} was used as the cost function for training teacher and student models in RemixIT.
We used the mean squared error between the estimated speech signal and bootstrapped mixture as $\mathcal{L}_{\rm Re2Re}$. For Re2Re\_reg, we set $\beta=100$ according to the development set.
We calculated DNS-MOS \cite{DNSMOS} scores on the 1-spk subset of the CHiME-5 dataset and SI-SDR [dB] on the reverberant LibriCHiME-5 dataset. 
More details about the datasets and baseline system are available in \cite{UDASE, UDASE_web}.

\subsection{Experimental results}
\begin{table}[t!]
\centering
\caption{SI-SDR [dB] in reverberant CHiME-5 dataset and DNS-MOS in 1-spk subset of CHiME-5 dataset achieved by our best systems and systems submitted to CHiME-7 challenge, ranked based on SI-SDR scores. Scores of other systems are obtained from \cite{UDASE_web}. The presence of ``VAD" indicates the version of the CHiME-5 dataset used for training.} 
\label{tab:result3}
\setlength{\tabcolsep}{4pt}
{\renewcommand{\arraystretch}{0.9}%
\begin{tabular}{@{\extracolsep{2pt}}lcccc@{}}
\thline
 & &\multicolumn{3}{c}{DNS-MOS}  \\
 \cline{3-5}
\multirow{-2}{*}{Systems}  & \multicolumn{1}{c}{\multirow{-2}{*}{\begin{tabular}[c]{@{}c@{}}SI-SDR \\ {[}dB{]}\end{tabular}}}                        & \multicolumn{1}{l}{OVRL} & \multicolumn{1}{l}{BAK}  & \multicolumn{1}{l}{SIG}  \\
  \hline
NWPU and ByteAudio         & 13.0         & 3.07 & 3.93 & 3.39 \\
Sogang ISDS1         & 12.4                                & 2.90                     & 3.60                     & 3.39                     \\
RemixIT-VAD                                                                           & \multicolumn{1}{c}{10.1}            & 2.84                     & 3.62                     & 3.28                     \\
Conformer Metric GAN & \multicolumn{1}{c}{7.8}             & 3.40                     & 3.97                     & 3.76                     \\
Sudo rm-rf                    & \multicolumn{1}{c}{7.8}             & 2.88                     & 3.59                     & 3.33                     \\
Input                                                             & \multicolumn{1}{c}{6.6}             & 2.84                     & 2.92                     & 3.48            \\
\hline
Re2Re      & 12.41  & 2.85     & 3.42    & 3.35     \\
Re2Re-VAD  & 12.41  & 2.79  & 3.39   & 3.32 \\
\thline
\end{tabular}}
\end{table}

We first compared the proposed Re2Re and Re2Re\_reg with the baseline system of CHiME-7.
\reftab{result1} shows SI-SDRs [dB] on the reverberant LibriCHiME-5 dataset and DNS-MOS scores in the 1-spk subset of CHiME-5 dataset. All models were trained from the Sudo rm-rf checkpoint provided by CHiME-7. 
The two proposed methods outperformed the baseline method in terms of SI-SDR, regardless of whether VAD was applied to the training data. Re2Re, using the N2N loss only, achieved SI-SDR about 0.71 dB and 1.08 dB higher than RemixIT. 
However, no improvement was observed for DNS-MOS. One possible reason is that Re2Re only considered the reconstruction error of the speech signal, resulting in a less accurate estimation of background noise. 
\reftab{result2} summarizes the SI-SDR[dB] and its std for each subset averaged over ten teacher models. 
The two proposed methods achieved better and relatively stable performance in all cases. 
The models trained on data without and with VAD achieved SI-SDR improvements of 0.99 and 1.62 dB on the 2-spk and 0.58 and 0.85 dB on the 3-spk subsets, respectively, while the improvements on the 1-spk subset were limited to 0.29 dB and 0.23 dB.
This could be another reason for the lack of improvement in DNS-MOS.
The standard deviations were approximately halved when trained on data without VAD and slightly reduced when trained on data with VAD, indicating that the performance of the student model relative to the teacher could be stabilized by N2N loss, even just as a regularization.
We then compared our best systems to those submitted to the challenge, whose results are summarized in \reftab{result3}. 
The proposed methods achieved performance comparable to the system ranked second in the challenge regarding SI-SDR and baseline RemixIT regarding DNS-MOS.

\section{Conclusions}
\label{sec:conclusion}
The paper proposed applying N2N learning to SE domain adaptation. The proposed method, called Remixed2Remixed, uses a teacher-student architecture, where a teacher model is pre-trained with OOD data and then used to generate pseudo-noisy pair data, and a student model is trained by minimizing an N2N-based loss function.
Experimental results on the CHiME-7 UDASE task revealed that Re2Re outperformed RemixIT regarding SI-SDR with a more stable performance.

\bibliographystyle{IEEEbib}

\end{document}